\documentclass[showpacs,showkeys]{revtex4-1}
\usepackage{graphicx}
\usepackage{subfigure}
\usepackage{latexsym}  
\usepackage{amssymb}   
 
\begin{document}

\title{Stripe-tetragonal phase transition in the 2D Ising
        model with dipole interactions: Partition-function zeros approach}
\author{\firstname{Jacyana} S. M.  \surname{Fonseca}}
\email{jacyanasaraiva@usp.br}
\author{\firstname{Leandro} G. \surname{Rizzi}}
\email{lerizzi@usp.br}
\author{\firstname{Nelson} A. \surname{Alves}}
\email{alves@ffclrp.usp.br} 
\affiliation{Departamento de F\'{\i}sica, FFCLRP,
           Universidade de S\~ao Paulo,
           Avenida Bandeirantes, 3900. \\
           Ribeir\~ao Preto 14040-901, SP, Brazil.}

\begin{abstract}
   We have performed multicanonical simulations to study the critical behavior of the two-dimensional Ising 
model with dipole interactions. 
   This study concerns the thermodynamic phase transitions in the range of the interaction $\delta$ where the 
phase characterized by striped configurations of width $h = 1$ is observed.
   Controversial results obtained from local update algorithms have been reported for this region, including 
the claimed existence of a second-order phase transition line that becomes first order above a tricritical 
point located somewhere between $\delta =0.85$ and 1.
   Our analysis relies on the complex partition function zeros obtained with high statistics from multicanonical 
simulations.
   Finite size scaling relations for the leading partition function zeros yield critical exponents $\nu$ that 
are clearly consistent with a single second-order phase transition line, thus excluding such tricritical 
point in that region of the phase diagram. 
   This conclusion is further supported by analysis of the specific heat and susceptibility of the orientational 
order parameter.
\end{abstract}

\keywords{Dipolar Ising model, complex partition function zeros, 
multicanonical simulations, striped phase, tetragonal phase}
\pacs{05.50.+q, 05.70.Fh,75.10.-b, 75.70.Kw, 75.40.Mg, 75.40.Cx}


\maketitle

\section{Introduction}

  The two-dimensional (2D) Ising model with nearest neighbor ferromagnetic exchange interaction
($J>0$) and dipolar interaction ($g>0$) presents a rich phase diagram because of these competing interactions.
  This model has been the focus of considerable theoretical interest, and the study of its phase diagram has 
revealed a variety of unusual magnetic properties.
   Moreover, at atomic level, it may give some insight into the interactions that form the striped phases 
observed in a number of ultrathin magnetic films \cite{debell_72_2000,portmann_422_2003} as a consequence 
of the reorientation transition of their spins at finite temperatures.
   The thermodynamic behavior has been investigated by analytical methods and Monte Carlo (MC) simulations, 
aiming at the determination of its critical behavior as a function of the ratio between the exchange and 
the dipolar interaction parameters, $\delta = J/g$.
   The Hamiltonian of this model is written as
\begin{equation}
          {\cal H} = -\delta \sum_{<i,j>} \sigma_{i} \sigma_{j} +  
                      \sum_{i < j} \frac{\sigma_{i} \sigma_{j}}{r_{ij}^{3}} \, .  
											                                                \label{eq:hamiltonian}
\end{equation}
  The variables $\sigma_i = \pm 1$ stand for the Ising spins in square lattices $L \times L$
and are supposed to be aligned out of plane.
  Here, we have adopted the convention \cite{cannas_D168_2002} of summing up over all distinct pairs of 
lattice spins at distances $r_{ij}$ to define the dipolar interaction $g$. 
  The distances $r_{ij}$ are measured in units of lattice.

   Analytical methods include some approximations like spin-wave theory and mean-field, 
\cite{kashuba-1993,abanov_54_1995,macisaac_51_1995,grousson-2000,biskup_2007,rastelli_PRB76_2007,giuliani_2006,
giuliani_2007,cannas_75_2007,giuliani_2011}
but conclusions like the fact that the spontaneous magnetization is zero for all temperatures
and that the $T=0$ configurations present patterns classified as regular checkerboards, irregular checkerboards, 
or stripes of different widths are important.
   The checkerboard pattern corresponds to the formation of alternate magnetic domains represented
by black and white rectangles.
   Each of these rectangles contain sites with identical spins and are denoted by \mbox{$<m,n>$}, where 
$m$ and $n$ stand for lattice units \cite{rastelli_PRB76_2007}. 
   Regular and irregular checkerboards are defined for $m=n$ and $m \neq n$, respectively.
   The striped patterns correspond to the formation of magnetic domains displayed in rectangles
of size \mbox{$<m,n>$} but with $n \rightarrow \infty$.

\begin{figure}[b!]
\begin{center}
\includegraphics[angle=0,width=0.5\textwidth]{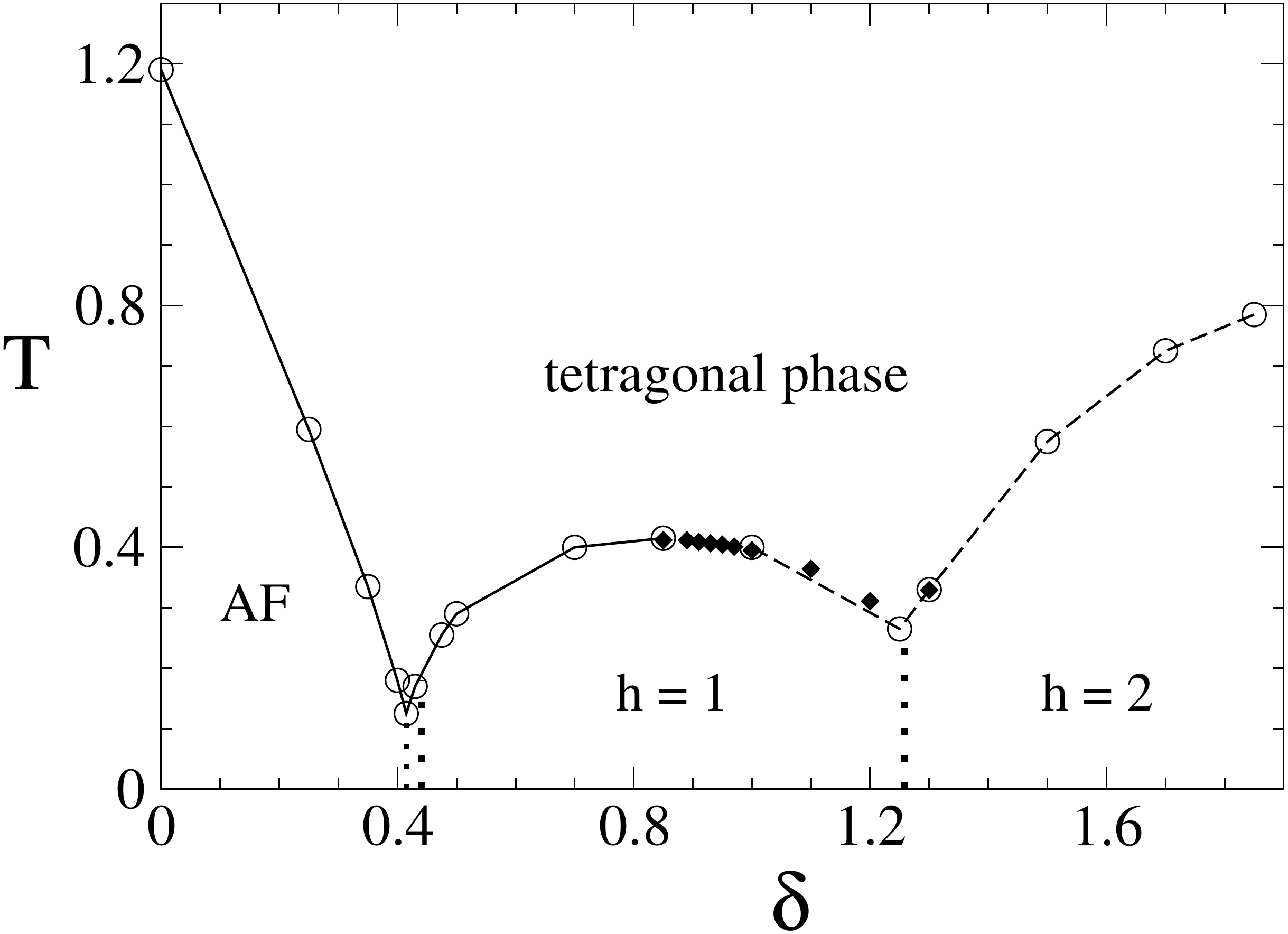}
\end{center}
\caption{Phase diagram: ($\circ$) data from Ref. \cite{rastelli_PRB76_2007},
($\blacklozenge$) this work. Vertical dotted lines represent the phase boundaries through a sequence
of phases characterized by antiferromagnetic (AF) ground state configurations, the $h=1$ 
and $h=2$ striped phases. The continuous line corresponds to the expected second-order transition except
in the narrow $\delta$ range $[0.4152, 0.4403]$, and the dashed line $(---)$ refers to a first-order one according 
to Ref.  \cite{rastelli_PRB76_2007} and  \cite{rastelli_73_2006}.}
\label{fig:phasedia}
\end{figure}

\begin{figure}[b!]
 \begin{center}
 \includegraphics[angle=0,width=0.55\textwidth]{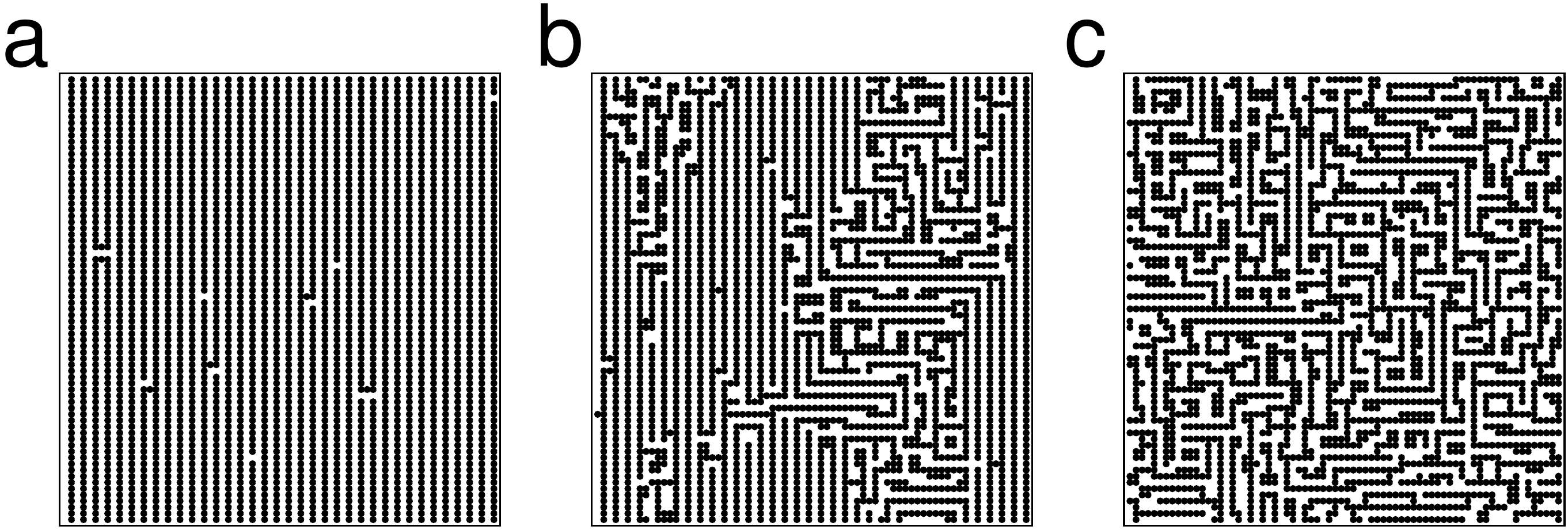}
 \end{center}
\caption{Spin configurations for $\delta=1.20$ and $L = 72$. 
(a)  striped phase: 
     $T=0.270$, $E/N=-0.4638$, $O_{hv}=0.9869$; 
(b) transition temperature:
    $T_c=0.311$, $E/N=-0.4096$, $O_{hv}=0.5039$;
(c) tetragonal phase:
    $T=0.350$, $E/N=-0.3539$, $O_{hv}=0.0476$.}
\label{fig:config120}
\end{figure}

\begin{figure}[b!]
 \begin{center}
 \includegraphics[angle=0,width=0.55\textwidth]{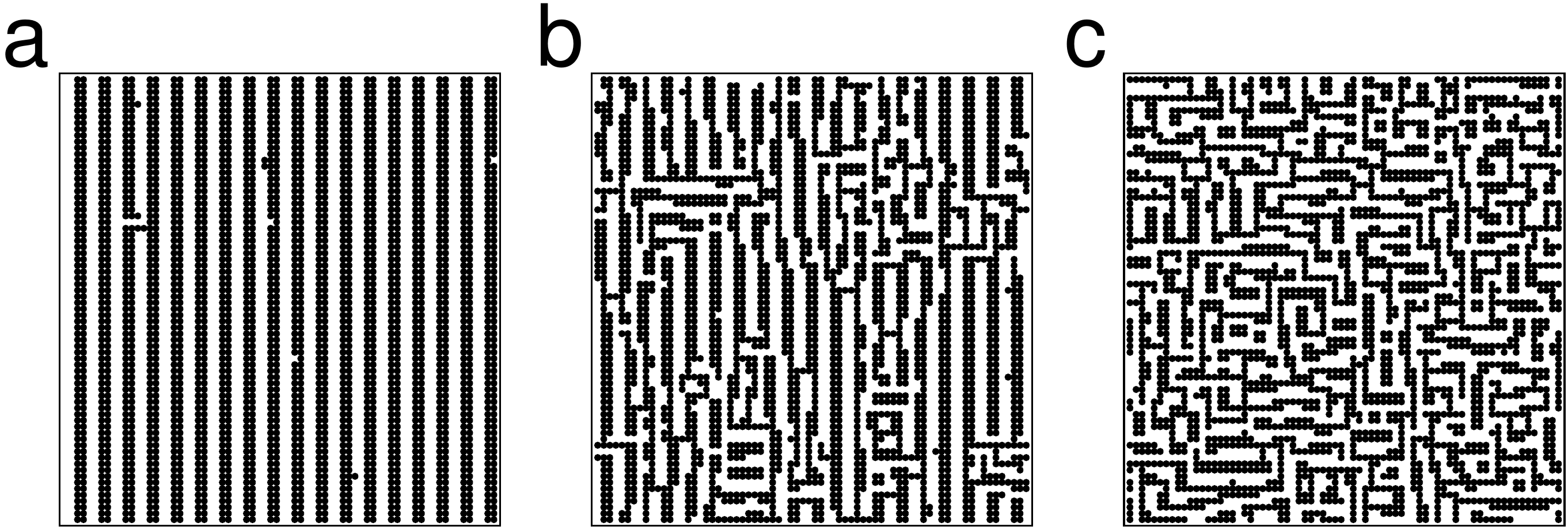}
 \end{center}
\caption{Spin configurations for $\delta=1.30$ and $L = 72$.
(a) striped phase:
    $T=0.290$, $E/N=-0.5058$, $O_{hv}=0.9831$;
(b) transition temperature:
    $T_c=0.329$, $E/N=-0.4397$, $O_{hv}=0.5173$; 
(c) tetragonal phase:
    $T=0.380$, $E/N=-0.3821$, $O_{hv}=0.0370$.}
\label{fig:config130}
\end{figure}

   Efforts have been made toward a rigorous theoretical proof for the spontaneous formation of these $T=0$ configurations \cite{giuliani_2006,giuliani_2007}.
	 The formation of such patterns as a consequence of the long-range character of the dipolar interaction
has been confirmed by MC simulations in different regions of the phase diagram $(\delta,T)$.
    In Fig. \ref{fig:phasedia} we show the phase diagram obtained from MC simulations for the $\delta$ range 
$[ 0, 1.9]$ where the above described ground-state patterns occur.
   The particular case $\delta=0$, a pure dipole interaction model, presents a continuous phase
transition with critical exponents in agreement with the ones in the universality
class of the 2D Ising model \cite{rastelli_73_2006,macisaac_46_1992}.
   For $0 <\delta < 0.4152$, the model presents antiferromagnetic (AF) ground-states characterized 
by stable regular checkerboard-like spin configurations \mbox{$<1,1>$}.
   Estimates from the specific heat indicate a continuous thermodynamic phase transition
associated with the change from this AF phase to a phase with broken orientational order, the so-called  
tetragonal phase \cite{rastelli_PRB76_2007}.
  In this phase, the magnetic domains lose their common orientation and try to assume the lattice symmetry. 
	The regular checkerboard configurations change to irregular checkerboard-like 
configurations \mbox{$<1,n>$} in the narrow range $0.4152 <\delta < 0.4403$.
  In this $\delta$ range, a thermodynamic first-order phase transition seems to take place 
\cite{rastelli_PRB76_2007}.
   For larger $\delta$ values, the ground state changes to spin configurations characterized by magnetic 
domains displayed in stripes of alternating spins, whose stripe width $h$ increases with 
$\delta$ \cite{macisaac_51_1995,giuliani_2007}.  
	 Striped configurations of width $h=1$ and $h=2$ occur for 
$0.4403 < \delta < 1.2585$ and $1.2585 < \delta < 2.1724$, respectively.
  Figures \ref{fig:config120} and \ref{fig:config130} contain these magnetic patterns obtained from our simulations for the couplings 
$\delta=1.20$ and $\delta=1.30$, respectively.
	In figure \ref{fig:config120}(a), the low-temperature ($T=0.270$) simulation at $\delta=1.20$ presents stripes of width $h=1$.
	Our simulations indicate a transition from the striped to the tetragonal phase at $T_c=0.311$ (figure \ref{fig:config120}(b)).
  The tetragonal phase is depicted in Figure \ref{fig:config120}(c).
  Figure \ref{fig:config130} presents magnetic patterns from simulations performed at $\delta=1.30$, a region where
stripes of width $h=2$ occur.

  In addition to the striped and tetragonal phases, a new domain in the phase diagram has been reported
for $\delta=2$, the so-called nematic phase in analogy with liquid crystals.
   In the nematic phase the system still keeps its long-range orientational order
but loses the spatial order exponentially.
   This new domain has been studied by MC simulations, and it has been found between the 
striped and the tetragonal phases.
   In this case, we would have two thermodynamic phase transitions:  stripe-nematic and 
nematic-tetragonal transitions.
    This new phase is located in a region of the $(\delta,T)$ plane that gives origin 
to a bifurcation of the line that separates the $h=2$ and $h=3$ phases \cite{cannas_73_2006,rizzi_2010}.
 
   A convincing determination of the thermodynamic phase-transition order is still lacking even for
such small $h$ values.
   In fact, controversial results about the order of the thermodynamic phase transition
as a function of $\delta$ have been reported in the literature.
   In particular, some MC results concerning square lattices for $\delta$ between 0.2 and 2 exhibit a phase 
diagram with a second-order transition line \cite{cannas_D168_2002,cannas_68_2003} for the thermodynamic 
transition between the ordered phases and the tetragonal one.
   On the other hand, the transition line appears to be first order for a $\delta$ range corresponding to 
$h \geq 1$ \cite{rastelli_PRB76_2007}, with the remark that for $\delta=0.85$ a second-order phase 
transition takes place with exponents $d\nu= 2.0\pm 0.1$, $\alpha=0.09\pm 0.07$, 
and $\gamma=1.75\pm 0.05$ at the critical temperature $T_c = 0.41$ \cite{rastelli_73_2006}.
    As usual, $\nu, \alpha$, and $\gamma$ are the correlation length, specific heat, and
susceptibility exponents, respectively.
   For the interaction $\delta=1$, it seems to be first order at $T_c=0.40$ \cite{rastelli_PRB76_2007}
and to present only a weak first-order character at $T_c = 0.404$ \cite{cannas_73_2006}.
   The above cited results would lead to the existence of two critical lines separated by a tricritical 
point for $\delta$ somewhere between 0.85 and 1. \cite{cannas_75_2007}

   The controversial results are a consequence of the dipolar term, which produces
large autocorrelations in MC time series obtained with local update algorithms
\cite{cannas_68_2003,cannas_78_2008,rizzi_2010}.
   Moreover, simulations have also been hampered because large lattice size simulations 
are very CPU time consuming due to this term, frustrating any convincing finite size
scaling (FSS) analysis.
   In this paper, we perform extensive multicanonical simulations for determination of the character 
of the thermodynamic transition from the $h=1$ phase, and to address the existence of a tricritical 
point. 
   The multicanonical algorithm (MUCA) generates a 1D random walk in the energy space, diminishing  
the problem of overcoming free energy barriers.
   We carry out a comprehensive analysis of the character of the phase transition
for values of $\delta$ from 0.85 up to 1.30 by means of the complex partition function
zeros \cite{fisher-1965,itzykson}.
   Partition function zeros analysis in the complex temperature plane has been successfully applied 
to spin models \cite{salvador1987,alvesB1990,alvesB1991}, 
lattice gauge theories \cite{alvesPRL1990,alvesB1992},
and protein models \cite{alvesPRL2000,ferrite2002}.
   This procedure has allowed us to explore critical aspects by means of FSS relations
for the first complex zero, leading to a precise characterization of the phase transition line.
   The conclusions based on the partition function zeros are further supported by analysis of 
the specific heat and susceptibility of the orientational order parameter.
    Analysis of these thermodynamic quantities allows us to calculate the 
critical exponents $\alpha/\nu$ and $\gamma/\nu$.
   It is well known that the renormalization-group fixed point picture for $d$-dimensional systems 
in the $L^d$ block geometry characterizes a first-order phase transition by the particular value
of the critical exponent $d\nu=1$ \cite{fisher-nu,decker_1988}.
   This, in turn, gives $\alpha=1$ and $\gamma=1$ for a first-order phase transition, which produce
the expected dependence of the thermodynamic quantities on the volume $L^d$ and
has been supported in a number of Monte Carlo studies	\cite{fukugita,alvesB1991}.

   Our results rely on data collected from lattice sizes up to $L=72$. 
   We report precise estimates for the infinite volume critical temperatures and critical 
exponents $\nu$, $\alpha/\nu$, and $\gamma/\nu$ for values of $\delta$ from 0.85 up to 1.20. 
   We have included the interaction $\delta=1.30$ in the $h=2$ phase for comparative purposes.
   In Sec. \ref{sec:mucasim}, we briefly review the main aspects of MUCA, and the protocol 
devised for updating of the multicanonical parameters.
   In Sec. \ref{sec:results}, results from FSS relations for the first complex zero, specific heat, and susceptibility
are compiled, to produce the estimates for the critical exponents.
   The final Sec. \ref{sec:conclusions} presents a summary and our main conclusions.

\section{Multicanonical simulations}
\label{sec:mucasim}

   The multicanonical algorithm \cite{berg-fields,berg-2003}, like other generalized algorithms 
\cite{okamoto-2001}, significantly improves the sampling of configurations.
   This algorithm assigns a weight $w_{mu}(E) \simeq 1/n(E)$, where $n(E)$
is the density of states and $E={\cal H}(\{ \sigma_{i} \})$ is the energy of a state given by the 
spin configuration $\{ \sigma_{i} \}$, with $i=1,..., L^2$, as defined in Eq. (\ref{eq:hamiltonian}).
	 Therefore, the multicanonical method is expected to produce flat energy histograms 
$H_{mu}(E) \propto n(E) w_{mu}(E)$ under the following probability condition
\begin{equation}
 p(E \rightarrow E') =\min \left[ 1, \frac{w_{mu}(E')}{w_{mu}(E)} \right]   \label{prob_accpt}
\end{equation}
for sufficiently long simulation times.
  
	 The multicanonical weight $w_{mu}(E)$ is {\it a priori} unknown. 
   A numerical estimate of $w_{mu}(E)$ is usually obtained by considering the Boltzmann entropy 
$S(E)= \ln n(E)$ ($k_B=1$), and the following parameterization 
for the entropy $S(E)=b(E)E-a(E)$,  where $a(E)$ and $b(E)$ are called multicanonical parameters. 
	 Hence,  the multicanonical weight is given by $w_{mu}(E)={\rm exp}[-b(E)E+a(E)]$, 
with the parameter $a(E)$ related to a multicanonical free energy and $b(E)$ related 
to the inverse of the microcanonical temperature.
  
   The implementation of MUCA requires energy discretization.
	 An integer label $m$ is introduced to facilitate our histogramming of energy data.
   This label defines energy bins of size $\varepsilon$, $E_m = E_0 + m\varepsilon$,
with $m = 0, \cdots, M$. 
   All the energies in the interval $[E_m, E_{m+1}[$ are in the $m${\it th} energy bin 
and contribute to the histogram $H_{mu}(E_m)$. 
   The constant $E_0$ is defined as a reference energy just below the ground-state energy.
   We have verified that $\varepsilon=1$ is a convenient discretization.
	
   The parameters $a(E)$ and $b(E)$ are estimated from  $N_{r}$ recursion steps.
	 Each step updates the multicanonical parameters through
the following equations \cite{berg-fields},
\begin{eqnarray}
 a^{n}(E_{m-1}) & = & a^{n}(E_{m}) + [b^{n}(E_{m-1})-b^{n}(E_{m})]E_{m}~,         \nonumber \\   
 b^{n}(E_{m})   & = & b^{n-1}(E_{m}) + 
   [ \ln \hat{H}^{n-1}_{mu}(E_{m+1}) -\ln \hat{H}^{n-1}_{mu}(E_{m}) ] / \varepsilon ~,    \label{recu}
\end{eqnarray}
where $n$ ($n=1, \cdots, N_r$) labels the recursion steps and $N_r$ amounts to how long 
this update procedure is enforced in order to obtain reliable estimates for  $w_{mu}(E)$.
   The choice $\hat{H}^{n}_{mu}(E_{m})=\max[h_{0},H^{n}_{mu}(E_{m})]$, 
with $0<h_{0}<1$ for all discretized energies, is a technical choice to avoid $H(E_m)=0$ \cite{berg-fields}.
   It is convenient to compute the above recurrence relations with the initial conditions 
$a^{0}(E_{m})=0$ and small values for $b^{0}(E_{m})$ if the simulation uses a hot-start initialization.
	 The $n${\it th} recursion step needs the calculation of $H_{mu}(E_{m})$ from the previous
weight $\{a^{n-1}, b^{n-1}\}$, obtained with $n_s$ MC sweeps.
   Usually, the number $N_r$ is defined {\it a posteriori} when the multicanonical parameters
present some convergence.
     
    To determine the multicanonical parameters, we have devised the following protocol.
	  Each recursion step is implemented after collection of $H_{mu}$ data
by sampling configurations between two extremal energies $E^*_{-}$ and $E^*_{+}$,
with $E^*_{-} < E^*_{+}$. 
    A round trip is defined as the number of sweeps necessary to go from configurations
with the lowest reference energy $E^*_{-}$ to the ones with a fixed high energy $E^*_{+}$ and back.
	  A round-trip walk may also start at any energy between $E^*_{-}$ and $E^*_{+}$. 
    The multicanonical update procedure Eq. (\ref{recu}) is performed with a variable number of
MC sweeps necessary for the attainment of three of such round trips.
    This number of round trips is chosen to ensure samplings across the energy
landscape in a reasonable simulation time.
    To avoid too long simulation time to achieve the next $(n+1)${\it th} multicanonical recursion, a fixed
number of sweeps $n_s(n)$ is set as the limiting number of MC updates. 
    Thus, new multicanonical parameters are obtained as soon as one of the following
conditions is observed: 
a) three round trips or b) a number of MC sweeps greater than three times the average
number of MC sweeps counted in the previous multicanonical simulations,
\begin{equation}
  n_s(n) = \frac{3}{n-1} \sum_{i=1}^{n-1} n_s(i) \, .            \label{eq:ns}
\end{equation}
    After each multicanonical update, $E^*_{-}$ is replaced with
the minimum energy among the sampled energies in the previous simulation. 
    This establishes a new (and larger) energy interval where new round trips must occur.
    This protocol helps us to keep a reasonable number of tunneling events even for 
large lattice sizes at the price of longer CPU times.
    A further improvement of the multicanonical weight $w_{mu}^{N_r}$ is achieved with an extra
MUCA update, which consists of $n_{MC}$ MC sweeps necessary for the performance of 20 round trips. 
    Table I lists only the number of sweeps $n_{MC}$ as a function of the lattice size $L$ 
for different interactions $\delta$ that are necessary for the accomplishment of this final update. 
  	With this final estimate of the multicanonical weight $w_{mu}(E)$, we proceed to data production.
		Our data production amounts to 16 independent energy time series, each one produced with $n_{MC}$ sweeps.
    Thus, the data analysis for the  smallest lattice size $L=12$ and $\delta=0.89$
relied on $\simeq 1.8\times 10^5$ measurements, while in the case of 
the largest lattice size $L=72$ and $\delta=1.30$ it amounted to $\simeq 1.62\times 10^8$ measurements.
    We can anticipate that the large number of measurements for $\delta=1.30$, compared with the smaller 
$\delta$ values, is related to the effort of overcoming the free energy barrier as a consequence of 
a first-order phase transition at this interaction.  
	 
  We have carried out simulations with periodic boundary conditions to minimize border effects.
  Thus, all distances $r_{ij}$  must include sites in the infinitely replicated simulation box in both directions. 
  This boundary condition adds an infinite sum over all images of the simulation box because of the dipole 
term in the Hamiltonian.
	The infinite sum was computed by means of the Ewald summation technique.
  This technique splits the infinite sum over all images of the system into two quickly converging sums, namely
the direct sum, which is evaluated in the real space, and the reciprocal sum, carried out in the reciprocal space, 
as well as a self-interaction correction term \cite{CompPhysComm.95.1996, JChemPhys.106.1997}. 
   We set the Ewald parameter $\alpha$ to 3.5 in all the simulations.
   This parameter determines the rate of convergence between the two sums.

		An important consequence of MUCA data production is the estimation of canonical averages 
of thermodynamic quantities $A$ over a wide range of temperatures $T=1/\beta$ by using the 
reweighting technique \cite{ferrenberg_61_1988}
\begin{equation}
\overline{A(\beta)} =
   \frac{\sum_{k} \,A_k\, [w_{mu}^{N_{r}}(E_{k})]^{-1} \exp{(-\beta E_k )} }
  	    {\sum_{k} [w_{mu}^{N_{r}}(E_{k})]^{-1} \exp{(-\beta E_k } )} \, .     \label{rewei}  
\end{equation}
	 This contrasts with the Metropolis algorithm, where the reweighting is restricted to a 
very narrow temperature range around the fixed MC simulation temperature.
   After reliable estimates for the MUCA weight, one can evaluate the density of
states
\begin{equation}
   n(E) = H_{mu}(E) w^{-1}_{mu}(E), 
\end{equation}
from which one can construct the partition function
\begin{equation}
 Z(\beta)  = \sum_E n(E) u^E,                              \label{eq:partition}
\end{equation}
where $u=e^{-\beta}$.
   The complex solutions in $u$, $\{{\rm Re}(u), {\rm Im}(u)\}$, describe the critical behavior of the system.
   These solutions correspond to the so-called Fisher zeros \cite{fisher-1965,itzykson}.

\section{Results}
\label{sec:results}
	
\subsection{Partition function zeros}	
	
    Let us consider the complex zeros of Eq. (\ref{eq:partition}) ordered according to 
their increasing imaginary part.
    For a sufficiently large lattice size $L$, the leading partition function zero 
$u_1^0(L)$ can be used to obtain the critical exponent $\nu$ through the FSS relation
\cite{itzykson}, 
\begin{equation}
  u_1^0(L) = u_c + A L^{-1/\nu}[1+O(L^{y})]~, ~~~~~~ y<0  .      \label{eq:r2}
\end{equation}
    This relation shows that the distance from the closest zero $u_1^0$ to the
infinite lattice critical point $u_c = e^{-\beta_c}$ on Re($u$) scales with the lattice size.
    If we disconsider finite size corrections, the exponent $\nu$ can be obtained from the linear regression
\begin{equation}
-\,{\rm ln}\, |u_1^0(L)-u_c| = \frac{1}{\nu}\,{\rm ln} (L) + a~. \label{eq:r3}
\end{equation}
    Since the exact critical temperature is unknown, and because the real part of $u$
presents weaker dependence on $L$ as compared to the imaginary part of $u$, it is
usual to replace $| u_1^0 - u_c|$ with its imaginary part, so as to avoid a multiparameter fit.

    With the discretization $\varepsilon$, Eq. (\ref{eq:partition}) becomes a polynomial
in $u$ and it can be solved with MATHEMATICA for $L \leq 32$. 
    Larger lattices present huge numbers for the density of states, which makes the
scan method in the complex temperature plane the only way of obtaining complex zeros 
\cite{alvesC1997}.
	  The leading complex zeros are presented in Tables II, III, and IV as a function of $L$
for different $\delta$ values.

  	Now, considering the real part of those zeros,
${\rm Re}\,[\beta_1^0(L)] =
 -1/2\, {\rm ln}\{ [{\rm Re}\,u_1^0(L)]^2\, + [{\rm Im}\,u_1^0(L)]^2 \}$, 
one can estimate the critical temperatures through the following FSS fit \cite{fukugita}:
\begin{equation}
 {\rm Re}\,[\beta_1^0(L)] = \beta_c^0 + b L^{-1/\nu} ~.     \label{eq:r4}
\end{equation}
   This fit yields the critical temperatures $T_c^0$ displayed in Table V,
where we have included the exponents $d\nu$.

\begin{figure}[b]
 \begin{center}
 \includegraphics[angle=0,width=0.5\textwidth]{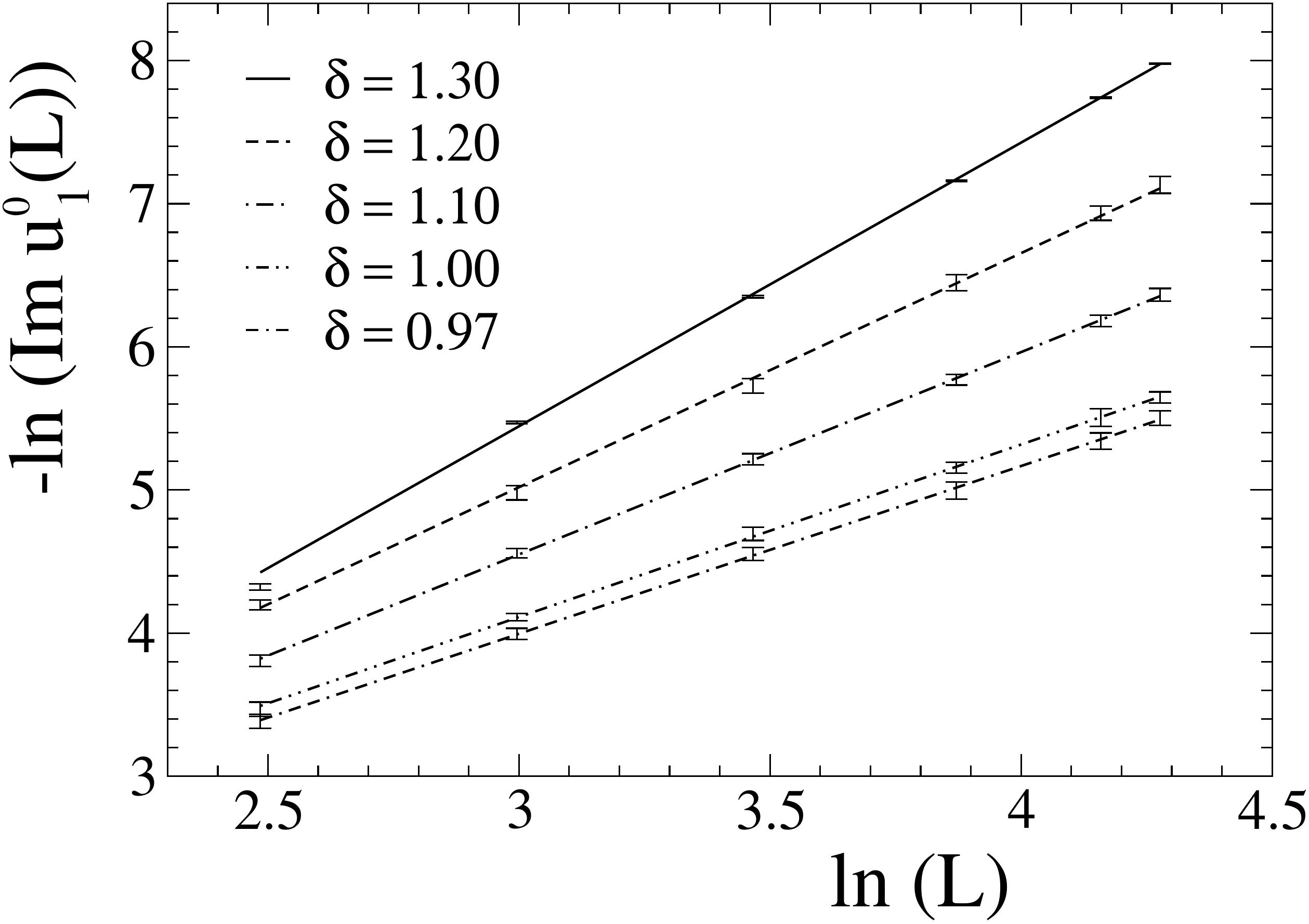}
 \end{center}
\caption{Finite-size scaling fits of the leading complex zeros for some $\delta$ couplings.}
\label{fig:linear}
\end{figure}

	 The quality of the linear fits can be stated in terms of the goodness-of-fit of
the model \cite{recipes}.
   The goodness-of-fit $Q$, $(0 \le Q \le 1)$ is related to $\chi^2$ and, 
as a general rule, if $Q$ is larger than 0.1, then the fit is believable.
   The linear fits for evaluation of $\nu$ present $Q$ as large as 0.98 for $\delta \leq 1.10$, $Q=0.70$ for
$\delta=1.20$, and a very small $Q$ value $\simeq 10^{-10}$ for $\delta =1.30$.
   Figure \ref{fig:linear} illustrates our linear fits for $\delta \geq 0.97$.
	 The data fit nicely, confirming the linear dependence on ${\rm ln}(L)$.
	 However, the very small $Q$ value for $\delta=1.30$ seems to be a consequence of high statistical precision
for these zeros (see Fig. \ref{fig:linear}), which reveals the presence of some systematic bias.
   It is known that corrections to FSS relations give a better fit for first-order phase transitions \cite{fukugita}.
   However, this would require larger lattice sizes for the attainment of reliable estimates from a multiparameter fit.
   The asymptotic behavior of $T_1^0(L)$ is determined with high $Q$ values for all $\delta$ interactions.	
	 Data collected in the third column of Table V shows a consistent trend toward $d\nu=1$ as we move on 
the critical line in the direction of higher $\delta$.
   The value $d\nu=1$ is only reached for the interaction in the $h=2$ phase.
	 Thus, these results clearly exclude first-order phase transitions from the $h=1$ phase.

		Results for $T_c^0$ are depicted in Fig. \ref{fig:phasedia} with the symbol ($\blacklozenge$).
		In this figure we also show the values obtained from Ref. \cite{rastelli_PRB76_2007} and, in particular,
we note that the values for $\delta=0.85$ and 1.3 are surprisingly good as compared to $T_c^0$, since they 
are obtained from a single lattice size $L=48$.

\subsection{Specific heat and susceptibility}

   To further characterize the order of the phase transitions, we have studied the specific heat,
\begin{equation}
    C_v(T) \ =\ \frac{1}{T^2 N} (\langle E^2 \rangle - \langle E\rangle^2) \, ,        \label{cv}
\end{equation}	
and the susceptibility
\begin{equation}
\chi(O_{hv}) = N \left( \langle O_{hv}^{2} \rangle  -  \langle O_{hv} \rangle^{2} \right),
\end{equation}
associated with the orientational order parameter \cite{ibooth_75_1995},
\begin{equation}
O_{hv} = \left| \frac{n_{v}-n_{h}}{n_{v}+n_{h}} \right|,
\end{equation}
over a (continuous) range of temperatures by reweighting MC data according to Eq. (\ref{rewei}).
  The quantities $n_h$ and $n_v$  are the number of horizontal and vertical bonds
of the nearest neighbor antiparallel spins, respectively.
  This order parameter is $+1$ in the striped ground state, and it vanishes at high temperatures
where orientational symmetry of the striped domain is broken.
	 A very common way of obtaining the critical exponents is through the FSS relations for
the maximum  of the specific heat
\begin{equation}
    C_{v}|_{\rm max}(T_c(L),L) \propto L^{\alpha/\nu} \, 
\end{equation}
and for the maximum of the susceptibility,
\begin{equation}
   \chi_{\rm max}(T_c(L),L) \propto L^{\gamma/\nu} \, ,
\end{equation} 
where $T_c(L)$ is the finite size critical point.
   Again, an FSS relation like Eq. (\ref{eq:r4}) is applied to the temperatures $T_c(L)$, where
the maxima of $C_v(T,L)$ and $\chi(T,L)$ occur, to yield the infinite volume critical temperature
$T_c^{C_v}$ and $T_c^{\chi}$, respectively.
   Table V summarizes these temperatures and the critical exponents for $C_v$ and $\chi$.
   The temperatures $T_c^{C_v}$ and $T_c^{\chi}$ are then evaluated with $\nu$
obtained from the hyperscaling relation $\alpha = 2 -d\nu$,  with data displayed in the 5{\it th} column of table V. 
   The goodness-of-fit of the linear fit for $C_v$ is about 0.5 for  $\delta \leq 1.20$.
	 Again, it decreases to a very small value $Q \simeq 10^{-5} $ for $\delta=1.30$.
	 The linear fit of $\chi$ presents $Q \simeq 0.8$ for $\delta \leq 1.20$ and also decreases
to $10^{-5}$ for $\delta=1.30$.

\begin{figure}[t!]
 \begin{center}
 \includegraphics[angle=0,width=0.44\textwidth]{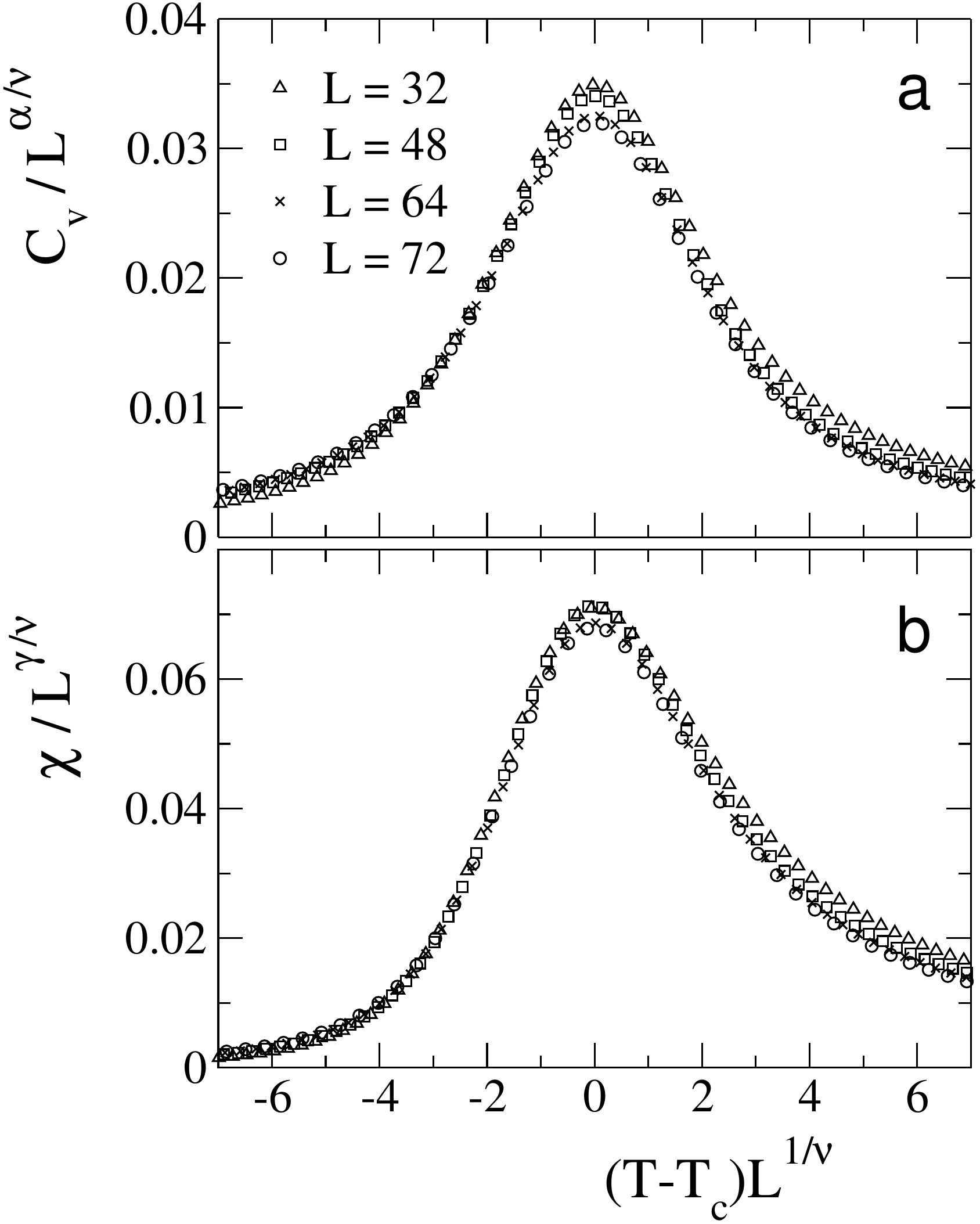}
 \end{center}
\caption{Finite-size scaling plots for the (a) specific heat and (b) susceptibility as a function
of the temperature for $\delta=1.20$.}
\label{fig:cv120}
\end{figure}

\begin{figure}[b!]
 \begin{center}
 \includegraphics[angle=0,width=0.44\textwidth]{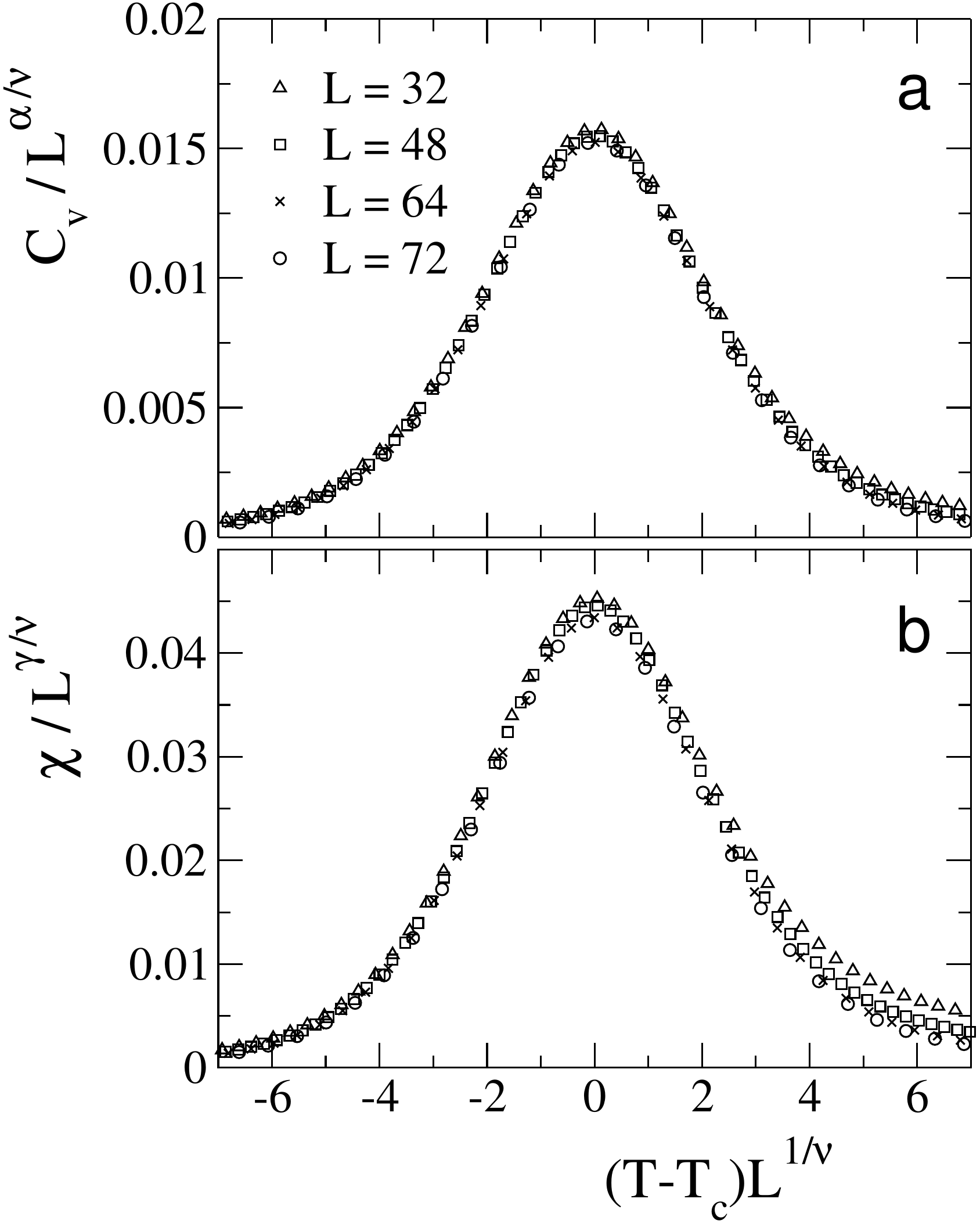}
 \end{center}
\caption{Finite-size scaling plots for the (a) specific heat and (b) susceptibility as a function
of the temperature for $\delta=1.30$.}
\label{fig:cv130}
\end{figure}

		The critical exponents $\alpha/\nu$ in the 5{\it th} column (Table V) clearly exclude any possibility 
of a first-order phase transition from the $h=1$ phase, while it strongly indicates this possibility at $\delta=1.30$.
    The statistical error bar exclude the value $\alpha/\nu =2$ at $\delta=1.30$, but the small $Q$ value may 
indicate the presence of systematic bias.
		The results from the susceptibility are less prompt to make satisfactory claims about the order
of the phase transition only at $\delta=1.2$ and $1.3$.
    Again, those results may be due to the missed corrections to the FSS relation, as expected at first-order 
phase transitions.
  	Figures \ref{fig:cv120} and \ref{fig:cv130} display the FSS plots for $C_v(T_c(L),L))$ and $\chi(T_c(L),L))$ for 
$\delta=1.20$ and $1.30$, respectively.
    These figures have helped us observe how satisfactory the FSS are.
	
\section{Summary and Conclusions}
\label{sec:conclusions}
    
		We have performed analysis of the complex partition function zeros from multicanonical simulations.
		The sampling with this algorithm is known to be efficient when it comes overcoming the free-energy barrier problem 
in simulations of complex systems as compared to the usual local update algorithms.
    MUCA is based on a non-Boltzmann weight factor and performs a free one-dimensional random 
walk in the energy space.
		A protocol has been devised for the determination of the multicanonical weight factor by ensuring that
enough measurements in the energy space are obtained.
    Moreover, by keeping such control over the number of round trips between the low and high-energy configurations,
we were able to determine the number of sweeps that is necessary for exploration of the energy space even for large lattice sizes.
	
    By using FSS relations involving the partition-function zeros obtained with high statistics, precise estimates 
for the infinite-volume critical temperatures and critical exponents $\nu$ were found for interactions $\delta$ 
corresponding to the $h=1$ phase.
    We also included an interaction ($\delta=1.30$) that produces stripes of width $h=2$ for comparative purposes.
    Analysis of the specific heat and susceptibility of the order-disordered parameter $O_{hv}$ gives further 
support for a second-order transition critical line from the $h=1$ phase.
    Infinite volume critical temperatures obtained from $u_1^0$, maximum of $C_v$ and 
$\chi$, are in full agreement, which help us draw a reliable part of the phase diagram $(\delta,T)$.
	  In conclusion, the study conducted with many $\delta$ interactions strongly indicates the existence
of a second-order critical line between the high-temperature tetragonal phase and the low-temperature ordered 
phase characterized by $h=1$. 
    The first-order character is found in our study only for the interaction $\delta=1.30$.
		This suggests that the second-order critical line ends at $\delta= 1.2585$, and that it becomes first-order
beyond this point.
    We assume that both transition lines are separated by a tricritical point at $\delta= 1.2585$,
because this point produces a line separating the $h=1$ and $h=2$ phases, and it bifurcates for generation of
the tetragonal phase.
    The precise temperature where this bifurcation happens has not been evaluated in the literature to the
best of our knowledge.

\section*{Acknowledgments}

    The authors acknowledge support by the Brazilian agencies FAPESP, CAPES, and CNPq.

\newpage

\begin{table}[ht]
\begin{center}
\small\addtolength{\tabcolsep}{-2pt}
\begin{tabular}{lr r r r r r r r r r }
\hline \\
[-0.35cm]
\hline\\
[-0.3cm]
$L${\large$\backslash$}$\delta$&0.85~~~&0.89~~~&0.91~~~&0.93~~~&0.95~~~&0.97~~~&1.00~~~  & 1.10~~~    & 1.20~~~    &1.30~~~      \\
\hline
$12$&15\,018   &~  11\,632&~    12\,032&~    14\,026&~    10\,186&~    16\,893&~    12\,279&~    24\,216&~    51\,171&~     73\,362 \\
$20$&65\,278   &~  47\,290&~    52\,098&~    46\,969&~    50\,127&~    78\,380&~    63\,333&~    91\,158&~   226\,934&~    269\,598 \\
$32$&181\,069  &~ 188\,620&~   149\,412&~   105\,008&~   158\,214&~   213\,036&~   179\,126&~   236\,436&~   959\,554&~ 1\,354\,207 \\
$48$&401\,195  &~ 483\,287&~   380\,451&~   645\,854&~   503\,469&~   484\,012&~   582\,030&~   879\,740&~2\,285\,886&~ 3\,256\,588 \\
$64$&1\,028\,097&~824\,022&~   938\,475&~   751\,156&~1\,104\,830&~1\,300\,445&~1\,296\,208&~2\,281\,061&~5\,579\,023&~ 8\,159\,701 \\
$72$&1\,283\,737&~1\,345\,861&~1\,421\,650&~1\,332\,818&~1\,709\,722&~1\,590\,101&~1\,697\,157&~1\,975\,543&~5\,682\,748&~10\,171\,715 \\
\hline \\
[-0.35cm]
\hline
\end{tabular}
\renewcommand{\tablename}{Table}
\caption{
 Number of MC sweeps $n_{MC}$ as a function of the lattice size $L$ for different interactions $\delta$.
 These numbers correspond to 20 round trips observed in the final MUCA update.}
\label{tab:n_mc}
\end{center}
\end{table}


\begin{table}[ht]
\begin{center}
\begin{tabular}{lllllll}\\
\hline \\
[-0.35cm]
\hline\\
[-0.3cm]
    &~~~~~~$\delta=0.89$&        &~~~~~~~$\delta=0.91$&         &~~~~~~~$\delta=0.93$&       \\
    &             &              &               &              &              &             \\ 
[-0.35cm]
$L$ &~~~~Re$(u_1^0)$&~~~Im$(u_1^0)$&~~~~~Re$(u_1^0)$&~~~Im$(u_1^0)$&~~~~~Re$(u_1^0)$&~~~Im$(u_1^0)$ \\
\\[-0.3cm]
\hline
\\[-0.3cm]
 12 &~ 0.0913(37)  & 0.0401(28)   &~~ 0.0923(36)  & 0.0381(29)   &~~ 0.0907(34)  & 0.0377(21)  \\
 20 &~ 0.0911(22)  & 0.0222(14)   &~~ 0.0909(24)  & 0.02117(95)  &~~ 0.0899(14)  & 0.0205(13)  \\
 32 &~ 0.0908(12)  & 0.01330(55)  &~~ 0.08967(96) & 0.01300(90)  &~~ 0.0886(11)  & 0.01237(62) \\
 48 &~ 0.09035(67) & 0.00842(57)  &~~ 0.08869(99) & 0.00807(51)  &~~ 0.08776(49) & 0.00770(50) \\ 
 64 &~ 0.08932(99) & 0.00615(39)  &~~ 0.08864(64) & 0.00598(32)  &~~ 0.08729(54) & 0.00568(39) \\ 
 72 &~ 0.08970(60) & 0.00556(39)  &~~ 0.08857(85) & 0.00522(41)  &~~ 0.08711(64) & 0.00502(31)\\
\hline \\
[-0.35cm]
\hline
\end{tabular}
\renewcommand{\tablename}{Table}
\caption{
    Complex partition function zeros for $\delta=0.89, 0.91$, and $0.93$.}
\end{center}
\end{table}


\begin{table}[ht]
\begin{center}
\begin{tabular}{lllllll}\\
\hline \\
[-0.35cm]
\hline\\
[-0.3cm]
    &~~~~~~$\delta=0.95$&        &~~~~~~~$\delta=0.97$&         &~~~~~~~$\delta=1.00$&       \\
    &             &              &               &              &              &             \\ 
[-0.35cm]
$L$ &~~~~Re$(u_1^0)$&~~~Im$(u_1^0)$&~~~~~Re$(u_1^0)$&~~~Im$(u_1^0)$&~~~~~Re$(u_1^0)$&~~~Im$(u_1^0)$ \\
\\[-0.3cm]
\hline
\\[-0.3cm]
 12 &~ 0.0895(31)  & 0.0358(25)   &~~ 0.0871(21)  &  0.0339(16)   &~~ 0.0844(18)  & 0.0312(16)  \\
 20 &~ 0.0886(14)  & 0.0194(10)   &~~ 0.0858(11)  &  0.01841(71)  &~~ 0.08298(86) & 0.01637(41) \\
 32 &~ 0.08708(82) & 0.01157(76)  &~~ 0.08497(55) &  0.01053(47)  &~~ 0.08162(62) & 0.00916(42) \\ 
 48 &~ 0.08634(51) & 0.00728(34)  &~~ 0.08446(42) &  0.00677(41)  &~~ 0.08124(43) & 0.00577(23)  \\
 64 &~ 0.08580(27) & 0.00513(17)  &~~ 0.08408(28) &  0.00479(28)  &~~ 0.08096(17) & 0.00406(25)  \\
 72 &~ 0.08583(31) & 0.00460(25)  &~~ 0.08391(26) &  0.00408(21)  &~~ 0.08068(15) & 0.00353(14)  \\
\hline \\
[-0.35cm]
\hline
\end{tabular}
\renewcommand{\tablename}{Table}
\caption{
Complex partition function zeros for $\delta=0.95, 0.97$, and $1.00$.}
\end{center}
\end{table}


\begin{table}[ht]
\begin{center}
\begin{tabular}{lllllll}\\
\hline \\
[-0.35cm]
\hline\\
[-0.3cm]
    &~~~~~~$\delta=1.10$&        &~~~~~~~$\delta=1.20$&         &~~~~~~~$\delta=1.30$&       \\
    &             &              &               &              &               &             \\ 
[-0.35cm]
$L$ &~~~~Re$(u_1^0)$&~~~Im$(u_1^0)$&~~~~~Re$(u_1^0)$&~~~Im$(u_1^0)$&~~~~~Re$(u_1^0)$&~~~Im$(u_1^0)$ \\
\\[-0.3cm]
\hline
\\[-0.3cm]
 12 &~ 0.0666(14)  & 0.02223(87)  &~~ 0.0365(14)  & 0.01504(51)  &~~ 0.0515(10)  & 0.01329(29)  \\
 20 &~ 0.06579(63) & 0.01049(34)  &~~ 0.03968(82) & 0.00687(35)  &~~ 0.05091(45) & 0.004203(34) \\
 32 &~ 0.06524(48) & 0.00544(21)  &~~ 0.04006(22) & 0.00326(17)  &~~ 0.04876(14) & 0.001745(15) \\ 
 48 &~ 0.06504(21) & 0.00312(12)  &~~ 0.040060(91)& 0.001584(91) &~~ 0.048100(66)& 0.0007782(50) \\
 64 &~ 0.06488(10) & 0.002070(83) &~~ 0.039935(59)& 0.000974(49) &~~ 0.047899(44)& 0.0004346(18) \\
 72 &~ 0.06476(11) & 0.0017227(77)&~~ 0.039934(39)& 0.000800(47) &~~ 0.047868(29)& 0.0003433(11) \\
\hline \\
[-0.35cm]
\hline
\end{tabular}
\renewcommand{\tablename}{Table}
\caption{
Complex partition function zeros for $\delta=1.10, 1.20$, and $1.30$.}
\end{center}
\end{table}


\begin{table}
\begin{tabular}{lllllll}
\hline \\
[-0.35cm]
\hline\\
[-0.3cm]
$~~\delta$&~~~~~~$T_c^0$   &~~~~$d\nu$  &~~~~~~~$T_c^{C_v}$ &~~~~$\alpha/\nu$ &~~~~~~~$T_c^{\chi}$ &$~~~~\gamma/\nu$ \\
\hline
$0.85$   &~~~0.41189(53) &~~1.837(76) &~~~~0.41240(48) &~~0.344(16)    &~~~~0.41200(51)  &~~1.519(19)  \\    
$0.89$   &~~~0.41168(53) &~~1.807(70) &~~~~0.41104(62) &~~0.364(20)    &~~~~0.41100(48)  &~~1.531(27)  \\
$0.91$   &~~~0.40887(50) &~~1.817(68) &~~~~0.40992(16) &~~0.375(19)    &~~~~0.40964(17)  &~~1.538(22)  \\
$0.93$   &~~~0.40681(19) &~~1.779(61) &~~~~0.40685(46) &~~0.399(20)    &~~~~0.40682(45)  &~~1.561(24)  \\
$0.95$   &~~~0.40435(17) &~~1.741(53) &~~~~0.40475(12) &~~0.424(20)    &~~~~0.40475(18)  &~~1.552(24)  \\
$0.97$   &~~~0.40108(40) &~~1.706(46) &~~~~0.40124(45) &~~0.461(19)    &~~~~0.40130(31)  &~~1.575(20)  \\
$1.00$   &~~~0.39499(37) &~~1.659(37) &~~~~0.39521(33) &~~0.522(17)    &~~~~0.39527(29)  &~~1.590(23)  \\
$1.10$   &~~~0.36429(29) &~~1.415(25) &~~~~0.36441(19) &~~0.888(21)    &~~~~0.36456(15)  &~~1.736(21)  \\
$1.20$   &~~~0.31102(32) &~~1.223(21) &~~~~0.31126(65) &~~1.496(28)    &~~~~0.31073(40)  &~~1.987(29)  \\
$1.30$   &~~~0.32929(72) &~~1.0093(28)&~~~~0.32892(15) &~~2.0183(66)   &~~~~0.32885(14)  &~~2.3193(82) \\
\hline \\
[-0.35cm]
\hline
\end{tabular}
\caption{Critical temperatures and critical exponents from complex partition function zeros, specific heat $C_v$, and 
susceptibility $\chi(O_{hv})$.}
\end{table}

\end{document}